\documentclass[runningheads]{svmult}

\usepackage{psfig} 
\usepackage{makeidx}   
\usepackage{graphicx}  
\usepackage{subeqnar}  
\usepackage{multicol}  
\usepackage{physprbb}  
\makeindex             

\title*{Clustering and Proto-Clusters in the Early Universe}

\def\spose#1{\hbox to 0pt{#1\hss}}
\def\lta{\mathrel{\spose{\lower 3pt\hbox{$\mathchar"218$}}
     \raise 2.0pt\hbox{$\mathchar"13C$}}}
\def\gta{\mathrel{\spose{\lower 3pt\hbox{$\mathchar"218$}}
     \raise 2.0pt\hbox{$\mathchar"13E$}}}

\let\jnlstyle=\rm
\def\refjnl#1{{\jnlstyle#1}}

\def\aj{\refjnl{AJ}}                   
\def\apj{\refjnl{ApJ}}                 
\def\apjl{\refjnl{ApJ}}                
\def\aap{\refjnl{A\&A}}                
\def\mnras{\refjnl{MNRAS}}             
\def\nat{\refjnl{Nature}}              

\author{Huub R\"ottgering\inst{1} \and Carlos De Breuck \inst{2}  \and 
Emanuele Daddi \inst{2} \and Jaron Kurk \inst{3} \and  George
Miley\inst{1}  \and  Laura Pentericci\inst{4} \and Roderik
Overzier\inst{1} \and Bram Venemans\inst{1}}

\authorrunning{Huub R\"ottgering et al.}
\institute{Leiden Observatory, Niels Bohrweg 2, NL-2333 CA Leiden,
The Netherlands \and 
European Southern Observatory, Garching, M\"unich, Germany \and 
INAF, Osservatorio Astrofisico di Arcetri, Firenze, Italy \and 
Dipartimento di Fisica, Universit\`a degli Studi Roma Tre, Italy } 

\begin{document}

\maketitle

\begin{abstract}
The clustering properties of clusters, galaxies and AGN as a function
of redshift are briefly discussed. It appears that extremely red
objects at $z\sim 1$, and objects with $J-K > 1.7$ and photometric
redshifts $2 < z_{phot} < 4$ are highly clustered, indicating that a
majority of these objects constitutes the progenitors of nearby
ellipticals. Similarly clustered seem luminous radio galaxies at
$z\sim 1$, indicating that these objects comprise a short lived phase
in the lifetime of these red objects. The high level of clustering
furthermore suggests that distant powerful radio galaxies (e.g. $z>2$)
might be residing in the progenitors of nearby clusters --
proto-clusters. A number of observational projects targetting fields
with  distant radio galaxies, including studies of Ly$\alpha$ and
H$\alpha$ emitters, Lyman break galaxies and (sub)mm and X-ray
emitters, all confirm that such radio galaxies are indeed located in
proto-clusters. Estimates of the total mass of the proto-clusters 
are similar to the masses
of local clusters. If the total star formation rate which we 
estimate for the entire proto-clusters  is sustained up
to $z\sim 1$, the metals in the hot cluster gas of local clusters can
easily be accounted for.
\end{abstract}

\section{Introduction}
An important theme in astrophysics is to understand the nature of
clustering of galaxies, active galaxies and even clusters of
galaxies. Observationally, two different methods have been employed
to measure clustering and its evolution with redshift.

A first method is to measure the level of clustering in various
distinct populations of objects. A number of statistical methods
have been used to quantify clustering including counts in cells,
nearest-neighbor statistics and the two-point correlation
function. By introducing the ``bias parameter'', which basically
measures the clustering of light compared to the underlying
clustering of dark matter, the clustering of objects in the
``local'' universe can be directly linked to the measured
fluctuations in the cosmic microwave background. For the art of
understanding how galaxies and AGN form, an interesting way to
rephrase aspects of this problem is to understand the evolution of
this bias parameter for various classes of galaxies and AGN. In
this contribution we will consider measurements of clustering of a
number of different classes of galaxies and AGN and how these
evolve as a function of redshift. Briefly, some implication for
the understanding of the nature of these objects will be
discussed.

A second method is to actually find and study ensembles of galaxies
in the early Universe for which the case can be made that they
will evolve into present day clusters. This will not only
constrain the formation and evolution of clusters, but will also allow
detailed studies of how the environment in which galaxies are born
influences their subsequent evolution. To find such early
clusters,  ``proto-clusters'', we have a number of programmes to
observe fields centered on extremely luminous radio galaxies with
redshifts in the range $2<z<5$. In all the well studied fields, it
seems that there are  significant excesses of galaxies compared to
the field, suggesting masses of the proto-clusters in the range of
$10^{14} - 10^{15}$ M$_\odot$.

\section{Clustering of galaxies and AGN}

A well established way of characterizing the clustering of objects
is through measuring their spatial two-point correlation function
(e.g. Peebles 1980). \nocite{pee80}
In the usual case of a limited number of
objects with well determined distances for which clustering is
being assessed, only the correlation length $r_0$ can be
constrained. This parameter indicates the spatial scale at which
the chance of detecting an object at a distance from a given object
is a factor of two more than for  random distribution.
Often, as in the case of new imaging surveys, the distances to the
objects are not known and only the angular clustering can
be determined. If the redshift distribution of such a population
is known, the spatial correlation length can be derived using the
well known Limber's equation (e.g. Daddi et al. 2000 and references
therein).\nocite{dad00a}

A consideration of the correlation length for a number of samples at
different redshifts of galaxies and AGN can be found in Overzier et
al. (2003) and R\"ottgering et al. (2003). \nocite{ove03,rot03b} At
low redshift, the most clustered objects are clusters of galaxies,
followed by elliptical galaxies and spirals. Detailed
measurements of the local correlation functions 
have recently been carried out on the basis of the  2dF
survey (Hawkins et al. 2003). \nocite{haw03} In the last few years, 
a number of interesting results have been obtained which constrain the
correlation length of various types of objects at $z\sim 1$. 

A major achievement of the VLA was the production of the NVSS radio
catalogue, which now contains 2 million radio sources observed at 1.4
GHz (Condon et al. 1998). \nocite{con98} Recently, two groups have
measured the angular correlation function of NVSS sources (Overzier et
al. 2003, Blake and Wall 2002), \nocite{bla02} and determined the
correlation lengths using the redshift distribution from Dunlop and
Peacock (1990). \nocite{dun90} In their analysis Overzier et
al. concluded that the most luminous radio sources at $z\sim 1$ are
clustered at about a level approaching that of local clusters. This may 
indicate that these sources are preferentially located in the
progenitors of nearby massive clusters, while  the less luminous sources
would then be associated with a field population.

The availability of large area infrared surveys
made it possible to study the clustering of extremely red objects
(EROs). The definition of the color cut that distinguishes EROs from
the general population is often taken to be $R-K > 5 $. 
The main two reasons that these  objects are so red are  that (i) they
straddle the 4000 \AA\ break of a $z\sim 1$ elliptical, or that 
(ii) the dust in a starburst galaxy has absorbed most of the optical light
(see Cimatti et al. 2003 and references therein). \nocite{cim03}
Various clustering measurements for EROs gave correlation lengths
of similar magnitude as found for the luminous radio sources (Daddi
et al. 2000, Roche et al. 2002, Firth et al.
2002). \nocite{roc02,fir02,dad00a} 
Taking into account
that the lifetime of radio sources is relatively small ($\sim 10^7$
years), the inferred space density of both EROs and radio galaxies are
similar (Mohan et al. 2002; Willott et al.
2001). \nocite{wil01,moh02} 
This all has the interesting
implication that luminous radio sources and EROs might be similar
objects at a different stage of their evolution. 

Measurements of clustering of optical quasars at $z\sim 1$ has been
done to an unprecedented accuracy by the 2dF team (Croom et
al. 2001). \nocite{cro01} Interestingly, the correlation lengths that
are found ($\sim 4-5$ Mpc) are significantly smaller than that for the
powerful radio galaxies. This has the immediate implication that these
optical quasars can not evolve into or descent from luminous radio
sources. Optical quasars are therefore probably associated with a
field population of objects with modest mass black holes as supposed
to the clustered radio galaxies with very massive black holes.

With the availability of wide and deep surveys conducted with 10-m
class optical telescopes, large numbers of  Lyman break galaxies
(LBGs) have  been detected. With these samples of LBGs it was
found that $r_0$ was constant over the measured redshift range of
$3<z<5$ (Hamana et al. 2004). \nocite{ham04} 
This is in agreement with  hierarchical models of
galaxy formation that state that the highest redshift galaxies are
the very biased tracers of the underlying dark matter
distribution.

The FIRES survey (Franx
et al. 2000) is a very deep survey with the VLT that is designed
to obtain a sample of high redshift galaxies selected in the
rest-frame optical rather than the restframe UV, like the LBGs. 
\nocite{fra00a} For this programme, in total 100 hours
were spent on observing the Hubble Deep Field South, reaching
in each of the infrared bands $J$, $H$ and $K$, limiting magnitudes
of  26.0, 24.9, and 24.5, respectivily (Labb\'e et al. 2003).
\nocite{lab03} One of the interesting discoveries was the presence of
a population of galaxies with $J-K > 2.3$. Although these objects had
photometric redshifts $z>2$, in general there was only limited overlap
with the LBG population (Franx et al. 2003). \nocite{fra03} This is
because these objects are too faint in the optical to detect the Lyman
break according to the usual criteria and/or are too red. Although the
number of such galaxies is limited, an attempt to measure the
clustering was presented by Daddi et al. 2003. \nocite{dad03} Down to
$K> 24$, it was found that  the galaxies with a $2 < z_{phot} < 4$
and $J-K > 1.7$ had a correlation length
a factor of 2-3 larger than LBG galaxies. Although the statistics are
clearly somewhat limited, the objects with $J-K > 2.3$ seem to be even
more clustered. Models of the evolution of the clustering of
ellipticals by for example Kauffmann et al. (1999) \nocite{kau99}
predicted that the correlation length for local elliptical and their
progenitors should be at the same high level. Since this is what we
observe, this is an important indication that these red objects are
indeed the progenitors of local ellipticals. Further evidence for this
comes from considerations related to their number density, masses and
sizes (van Dokkum et al. 2004, submitted).

\section{Radio galaxies at the centers of proto-clusters}

As already noted, the powerful
radio galaxies at $z\sim 1$ seems to have large correlation lengths,
suggesting that they are located in progenitors of local
clusters. This suggests that proto-clusters could be found in fields
containing distant and powerful radio galaxies. There is significant
additional supporting evidence that radio galaxies can be associated with
proto-clusters, including (i) they are amongst the most massive
objects at high redshifts, (ii) they have large Ly$\alpha$ halo's
whose outer regions are consistent with originating from a cooling
flow, (iii) 20 \% have high rotation measures as measured with the
VLA, indicating the presence of a dense medium. We therefore started a
project to study fields centered at distant powerful radio galaxies
with the aim of finding proto-clusters at a range of
redshifts. A second aim was to study the various classes of galaxies
in these proto-clusters.

\section{The proto-cluster associated with 1138-262 at $z=2.2$}

The first radio galaxy for which we studied its associated
proto-cluster in great detail was 1138-262 at $z=2.2$ (Kurk et al. 2003
and references therein) \nocite{kur03b} For a number of reasons this
was among the very best objects to study. It is among the brightest
object both at K-band and at radio wavelength. It has the highest
rotation measure (6200 rad/m$^2$) of a well studied sample of 80 $z>2$
radio galaxies. The morphology both in the radio, optical and near
infrared is very clumpy, consistent with simulation of forming
brightest cluster galaxy in which many star forming regions are
merging together to ultimately form a massive galaxy. With a size of
almost 200 kpc, the very luminous Ly$\alpha$ halo represents a
significant reservoir of gas from which part of the stars that will
make up the final system can be formed. The first programme was to
carry out deep narrow band imaging to obtain candidate Ly$\alpha$
emitting galaxies. This resulted in 70 candidates. As compared to the
field the overdensity was derived to be at least a factor of two.
Subsequent spectroscopy 
resulted in 14 confirmed galaxies at the redshift of the radio
galaxy. Deep narrow band imaging in the infrared combined
with infrared spectroscopy resulted in the detection of 7 confirmed
H$\alpha$ emitting galaxies. The K-band imaging data combined with the
optical contained 44 galaxies with $I-K >  4.3$. Unfortunately, these
objects have no measured redshift -- this is notoriously difficult --,
but their surface density peaks at the location of the radio sources.
Finally, we looked at the X-ray emitting galaxies using a deep Chandra
image. On the basis of either spectroscopy or colour information, for
6 objects the case could be made that they are at the redshift of the
proto-cluster. The spatial distribution of the various classes 
is given in Figure 1. In Figure 2, the surface density of these
classes as a function of distance from the radio galaxy is given. An
important point apparent from these two figures, is that the
distribution of H$\alpha$ galaxies seems much more concentrated than
the Ly$\alpha$ emitting galaxies.  Further differences are that the
population of H$\alpha$ emitters on average have brighter K-band
magnitudes and as a whole have a lower velocity dispersion. We
interpreted this as the population of H$\alpha$ emitting galaxies
being older and dustier and further in the process of having their
orbits virialized in the ``proto-cluster'' potential. It seems that
the seeds of the morphology-density relation are already in place at
$z=2.2$.  Following the analysis of Steidel et al. (1998) \nocite{ste98} 
as they
carried out for the high ``redshift spike'' at $z=3.1$, 
the mass associated with the over density is in the range $10^{14} -
10^{15}
$$M_\odot$. And indeed this is what is expected for a progenitor
of a nearby massive cluster.

\begin{figure}[ht]
\centerline{\psfig{figure=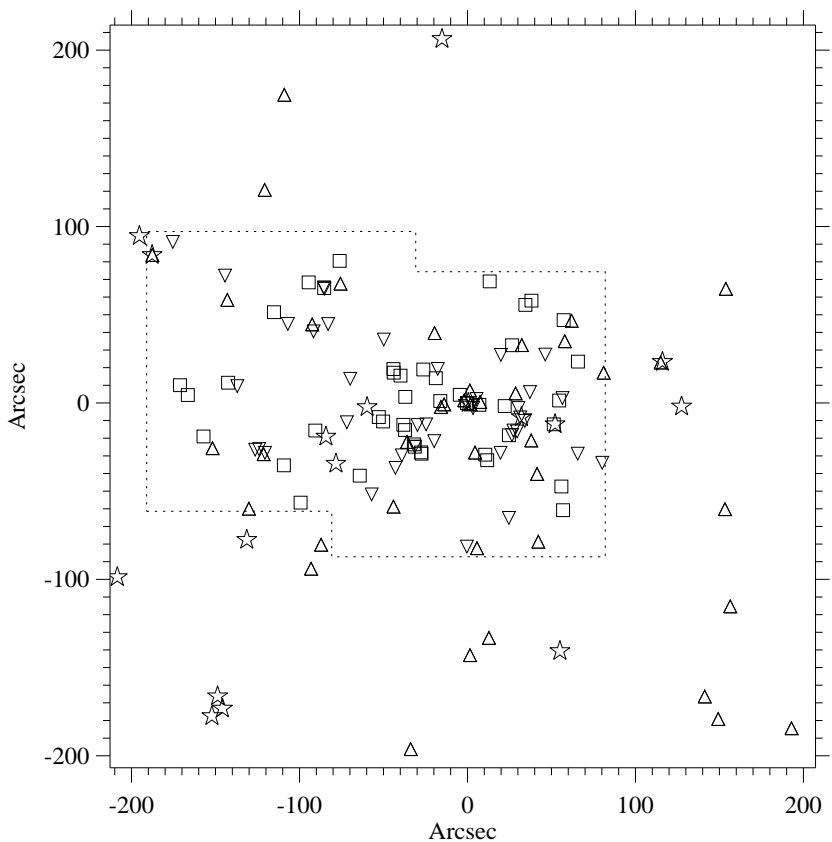,width=0.9\textwidth,clip=}}
\caption{ The spatial distribution of candidate cluster galaxies in
the field of the powerful radio galaxy 1138-262 ($z=2.2$).  Indicated
are the candidate H$\alpha$ emitters with downward pointing triangles,
candidate Ly$\alpha$ emitters with upward pointing triangles and extremely red
objects with squares. The X-ray emitters are indicated with stars.
The radio galaxy is located at the origin. The dotted
line indicates the boundaries of the two fields of the IR observations
with ISAAC/VLT within which H$\alpha$ emitters and EROs have been
searched for.  Note that spectroscopy yielded 15 confirmed Lya and 7
confirmed Ha emitters. }
\end{figure}

\begin{figure}[ht]
\centerline{\psfig{figure=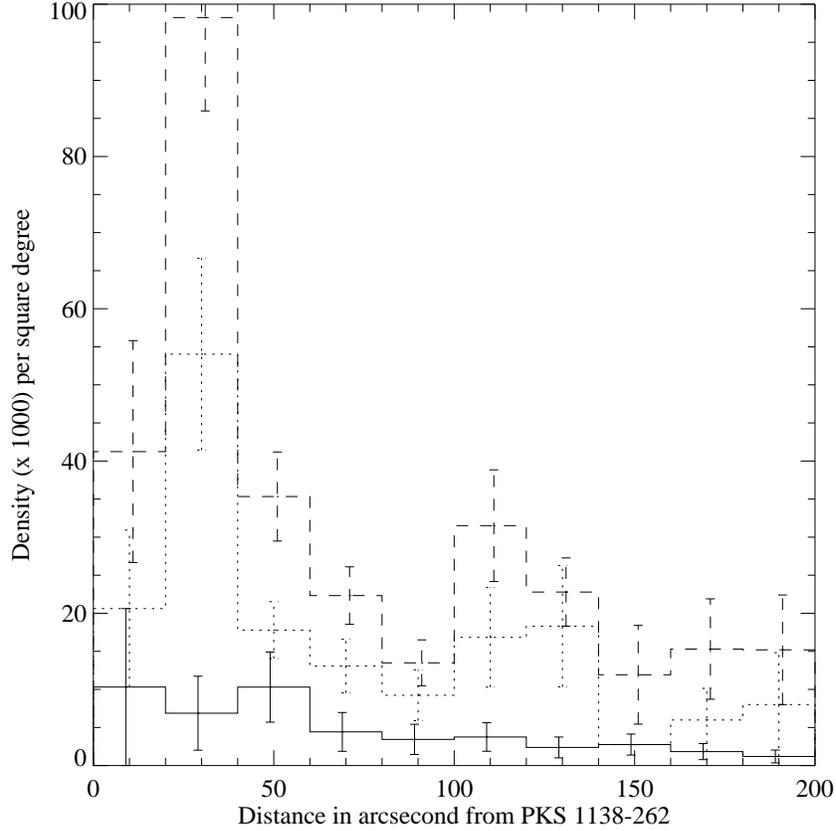,width=0.9\textwidth,clip=}}
\caption{The surface density of candidate cluster galaxies as a
function of distance from the powerful radio galaxy 1138-262
($z=2.2$).  Indicated are candidate Ly$\alpha$ emitters with black
lines,
candidate H$\alpha$ emitters with
dotted lines,  and extremely
red objects with dashed lines.} 
\end{figure}

\section{Towards a sample of proto-clusters}

The results on the proto-cluster of 1138-262  prompted us to
extend this  study to more objects. We initiated an ESO large
programme to search for Ly$\alpha$ emitting galaxies in potential 
proto-clusters around radio galaxies with redshifts in the range
$2<z<5$.  An account of the results of this programme has been
presented in this conference by Venemans et al.  The main conclusion is
that all of the 7 radio galaxies that have been studied in
sufficient detail show over densities and resulting masses similar
to that found for 1138-262.

\section{Hubble Space Telescope ACS imaging}

In the course of our ESO large programme, we studied the region of
the powerful radio galaxy 1338-193 at $z=4.1$ 
and subsequently found the highest redshift group of
galaxies. In total we have 50 candidate Ly$\alpha$ emitters
distributed over two partly overlapping $7\times 7$ arcminute field of
views of the FORS instrument of the VLT.  Of these, 32 are
spectroscopically confirmed. This field was studied in detail
using data from the new Advanced Camera for Surveys on the Hubble
space telescope (Miley et al.  2004). \nocite{mil04} It was observed in three
filters (F475m, F625m, F775m), for optimum sensitivity for
detecting LBG galaxies at $z\sim 4$.  Comparisons
with the surface densities in the Hubble Deep Field and the Subaru
deep field led us to conclude that there are a factor of 2 more
$z\sim 4$ LBGs  than in a random field. Given that the three
filters in principle select objects over the entire redshift range
of $3.5 < z < 4.5$, the actual spatial over density compared to the field
is likely to be well in excess of ten.

\section{mm/submm imaging with JCMT and IRAM}

Both the LB galaxies and the Ly$\alpha$ emitters are discovered at optical
wavelengths, which naturally biases the obtained samples to galaxies
that are relatively unobscured. With (sub)mm telescopes, a number of
galaxy fields have been observed with the aim of detecting very
dusty and potentially very obscured galaxies located in the
proto-cluster. Using the SCUBA array mounted on the JCMT, 7 radio
galaxy fields have been observed. An excess of dusty galaxies of about
a factor of two compared to the field has been found (Stevens et
al. 2003). \nocite{ste03} De Breuck et al. (2004, submitted) carried out a
detailed investigation of the field of the radio galaxy 1338$-$192 at
a redshift of $z=4.1$ using MAMBO  at the IRAM 30m. Combined with deep
optical and radio data 5 dusty objects were found with properties
consistent with being located at the distance of the proto-cluster.
This is further evidence for the reality of an excess of starburst 
galaxies in distant radio galaxy fields.

\section{Discussion}

Both from the analysis of the correlation function of galaxies and
AGN and the observational data, it seems that the evidence is now
very convincing that distant powerful radio galaxies are located in
``proto-clusters''. It is even plausible that every proto-cluster has
gone through a radio-active phase as it turns out that the space
density of powerful radio galaxies is comparable to the space
density of local clusters, taking into account the short 
lifetime of the radio activity ($10^6 - 10^7$ years). Again following the 
analysis of Steidel et al., the associated masses of the structures are 
in the range of 
$10^{14} - 10^{15}$ M$_\odot$.
A simple
estimate of the total star formation rate of all the galaxies
combined in the proto-clusters is in excess of 20,000 M$_\odot$ yr$^{-1}$. If 
such a high rate is sustained, then the high metal content in the
hot gas in $z<1$ clusters can be easily accounted for.


\section*{References} 

{Blake}, C. \& {Wall}, J. 2002, \mnras, 329, L37
\\
{Cimatti}, A., {Daddi}, E., {Cassata}, P., {et~al.} 2003, \aap, 412, L1
\\
{Condon}, J.~J., {Cotton}, W.~D., {Greisen}, E.~W., {et~al.} 1998, AJ, 115,
  1693
\\
{Croom}, S.~M., {Shanks}, T., {Boyle}, B.~J., {et~al.} 2001, \mnras, 325, 483
\\
{Daddi}, E., {Cimatti}, A., {Pozzetti}, L., {et~al.} 2000, \aap, 361, 535
\\
{Daddi}, E., {R{\" o}ttgering}, H.~J.~A., {Labb{\' e}}, I., {et~al.} 2003,
  \apj, 588, 50
\\
Dunlop, J. \& Peacock, J. 1990, \mnras, 247, 19
\\
{Firth}, A.~E., {Somerville}, R.~S., {McMahon}, R.~G., {et~al.} 2002, \mnras,
  332, 617
\\
{Franx}, M., {Labb{\' e}}, I., {Rudnick}, G., {et~al.} 2003, \apjl, 587, L79
\\
{Franx}, M., {Moorwood}, A., {Rix}, H., {et~al.} 2000, The Messenger, 99, 20
\\
{Hamana}, T., {Ouchi}, M., {Shimasaku}, K., {Kayo}, I., \& {Suto}, Y. 2004,
  \mnras, 347, 813
\\
{Hawkins}, E., {Maddox}, S., {Cole}, S., {et~al.} 2003, \mnras, 346, 78
\\
{Kauffmann}, G., {Colberg}, J.~.~M., {Diaferio}, A., \& {White}, S.~D.~M. 1999,
  \mnras, 307, 529
\\
{Kurk}, J., {R{\" o}ttgering}, H., {Pentericci}, L., {Miley}, G., \&
  {Overzier}, R. 2003, New Astronomy Review, 47, 339
\\
{Labb{\' e}}, I., {Franx}, M., {Rudnick}, G., {et~al.} 2003, \aj, 125, 1107
\\
{Miley}, G.~K., {Overzier}, R.~A., {Tsvetanov}, Z.~I., {et~al.} 2004, \nat,
  427, 47
\\
{Mohan}, N.~R., {Cimatti}, A., {R{\" o}ttgering}, H.~J.~A., {et~al.} 2002,
  \aap, 383, 440
\\
{Overzier}, R.~A., {R{\" o}ttgering}, H.~J.~A., {Rengelink}, R.~B., \&
  {Wilman}, R.~J. 2003, \aap, 405, 53
\\
{Peebles}, P.~J.~E. 1980, {The large-scale structure of the universe} (Research
  supported by the National Science Foundation.~Princeton, N.J., Princeton
  University Press, 1980.~435 p.)
\\
{R{\" o}ttgering}, H., {Daddi}, E., {Overzier}, R., \& {Wilman}, R. 2003, New
  Astronomy Review, 47, 309
\\
{Roche}, N.~D., {Almaini}, O., {Dunlop}, J., {Ivison}, R.~J., \& {Willott},
  C.~J. 2002, \mnras, 337, 1282
\\
{Stevens}, J.~A., {Ivison}, R.~J., {Dunlop}, J.~S., {et~al.} 2003, \nat, 425,
  264
\\
{Willott}, C.~J., {Rawlings}, S., \& {Blundell}, K.~M. 2001, \mnras, 324, 1
\\

\end{document}